\documentstyle[12pt]{article}

\begin{document}
\def
\dslash{\partial \!\!\! /}

\hfill{}

\vspace{7pt}
January 15, 1995\hspace{7cm}
\vspace{50pt}
\begin{center}
{\large\sc{\bf Double expansion in asymptotically free theories.}}

\baselineskip=12pt
\vspace{50pt}

B. Rosenstein$^ *$\,

\vspace{20pt}
Institute of Physics, Academia Sinica\,
Taipei, 11529\,
Taiwan, R.O.C.\,

\vspace{80pt}
\end{center}
\begin{abstract}
I propose an approximation scheme for asymptotically free
field theories combining both weak coupling and strong coupling
series.
The weak coupling expansion is used to integrate
the high frequency modes and
the resulting low energy effective theory is solved using the strong coupling
expansion. In some models there exists an intermediate scale at which both
expansions make sense. The method is tested on few low dimensional models for
which an exact solution is known.
\end{abstract}

\vspace{90pt}

*BARUCH@PHYS.SINICA.EDU.TW
\pagebreak
\vspace{1cm}

In strongly coupled theories like QCD there are plenty of qualitative
phenomenological models (quark, bag, string models, strong coupling
expansion ...),
but just one approximation which is under control: the asymptotic weak coupling
perturbation
 theory. This is only  applicable at energies
much higher than $\Lambda_{QCD}$.
It is commonly believed that for low energy physical quantities,
like hadron masses, there is no
small parameter in which one can expand. The theory
apparently does not contain a  natural small
parameter and
the expansion in a reasonable small "coupling" fails at
small energies.

Nevertheless, the relative success of
one of the "phenomenological" approaches  seems to indicate that it
contains at least part of the truth. This is the strong coupling
expansion which is usually performed using lattice regularization
with lattice spacing
 $\sim 1/({\rm few }\Lambda_{QCD})$ \cite{kreutz}.
Confinement and chiral symmetry breaking are easily seen
and the intricate spectrum is nicely accounted for,
at least as far as ordering and quantum numbers
are concerned, and even the mass ratios are reasonable \cite{kogut}.
At one time there was a hope that
one can extrapolate from the strong coupling series into the
"scaling region"
using Pade approximants \cite{kogut}. This would convert the method
from just a phenomenological model to a systematically improvable scheme.
Unfortunately, this method includes too much
guesswork to be controllable and it was impossible
to reach the scaling region from the apparent radius of convergence
of these series (about $\beta\sim 5$ in
$SU(3)$ Yang - Mills \cite{kogut}). As it is usually used, the strong coupling
series have very little to do
with the (asymptotically free) continuum limit we are
interested in. Therefore one should consider it just as a model for
 low energy phenomena
with the UV cutoff $M$ and the coupling $g$ which should
then be related to
$\Lambda_{QCD}$.

Even in much simpler asymptotically free theories the situation  is similar.
 In the d=2
$\sigma$ - model and the Gross - Neveu model strong coupling expansions were
performed \cite{elitzur} and  qualitatively
reproduce quite nicely the exactly
known \cite{zz} low energy features. Yet the continuum
limit is still unreachable. The available exact
S - matrices in these
models show what kind of nonanalyticity
in coupling constant is expected.
The general dependence of a physical quantity, for example
a S - matrix element,
on a physical coupling  $g$ has
the following structure \cite{davis}:
\begin{equation}
f(g)=\sum_{m=0}^\infty\sum_{n=0}^\infty a_{m,n}   g^m
e^{-n/g}\label{fGN}
\end{equation}
where $a_{m,n}$ are numerical coefficients.
It contains both polynomial and nonanalytic in $g$
(but analytic in $1/g$) "instantonic" exponential
factors. If one would be able to separate these
dependencies, by expressing $f$ as
\begin{equation}
f(g)=F(\alpha(g),\beta(g)),
\label{f}
\end{equation}
so that $F(\alpha,\beta)$ has a viable expansion in
 both variables, this would
provide a systematic method to approximate these theories. Here the
"weak coupling" $\alpha(g)$ is analytic in $g$ and  the "strong
coupling" $\beta(g)$ is analytic in
$1/g$.

In this paper I formulate such an attempt and test it on low dimensional
QFT like the just mentioned two. Of course,
one can wonder whether this is at all possible. {\it Apriori} it looks like
the domain of applicability  of the expansion in $\beta$ has a very little
chance to intersect with that of the expansion in $\alpha$. Indeed,
if $g$ is small,  $1/g$ is large and {\it vice versa}. Fortunately,
this does not always imply that for small
$\alpha(g)$, $\beta(g)$ should be large and {\it vice versa}.
Just a glance at strong coupling expansion results \cite{kogut}
shows that, although the asymptotic scaling region is out of reach, the
effective weak coupling near the strong coupling radius of convergence
is quite small. In the $SU(3)$ YM theory the radius of
convergence $\beta_{max}\equiv 6/g_{min}^2\sim 5$
corresponds (using naive perturbative RG) to  a relatively small effective
weak coupling
expansion parameter $\alpha\equiv g^2/4\pi\sim 0.1$. Even taking
into account  the fact that the $\alpha$ is a lattice one
(not the ${\overline {\rm MS}}$),
this corresponds to the values at which perturbation theory
for high momentum quantities is supposed to work. The same happens
in two dimensional asymptotically free theories. Consequently, there
is a (albeit
small) window in the coupling at which both $\alpha$ and $\beta$ are small
enough to produce a reasonable series. The "loop
factors" $1/(4\pi)^2$  in the practical weak coupling expansion
parameter $\alpha(g)$
are partly responsible for this. But this should be checked in any
particular case since, for example, the symmetry group factors tend to reduce
the window\footnote{The location of the "window" and even
 its existence may depend on the choice of coupling and therefore on
the regularization and the normalization scheme.}.

Now, I will outline the general method to calculate a physical quantity.
The model's lagrangian on the UV cutoff  $\Lambda$ scale is proportional to a
local dimension $d$ operator:
${\cal L}_{\Lambda}(\phi)=\frac {1}{g_\Lambda} {\cal L}_0(\phi)$,
where
 $g_\Lambda$ is a bare coupling constant and $\phi$ are  fields\footnote{ We
consider here only models  like pure
Yang - Mills theory or chirally invariant QCD without additional couplings.
The generalization to more complicated asymptotically
free theories is trivial.}.\newline

1. Aa a first step, one integrates the high frequency modes from $\Lambda$
down to  a certain scale $M$. $M$ should be larger than $\Lambda_{model}$
, so  that the weak coupling expansion
 (or the loop expansion)
 is applicable where $\Lambda_{model}$ is the corresponding dynamically
generated scale
of the "model".
The result is, of course, a  complicated functional: an effective action
of the type extensively studied recently \cite{polchinski}.
 It  typically allows the standard derivative expansion
 in powers of $1/M^2$ of the following form:
\begin{equation}
{\cal L}_M[\phi]=\sum_{i=-1}^\infty\sum_{j=0}^\infty
\tilde c_{ij} g_\Lambda^i \frac {1}{M^{2j}}
{\cal L}_j(\phi)
\label{L_Mgen}
\end{equation}
where ${\cal L}_j$ is a dimension $d+2j$ local operator.
The first operator  ${\cal L}_0$ is, of course, the original
 lagrangian now containing however the
low frequencies only. If the regularization is the lattice one,
it is simply a
 coarse grained version of the original lagrangian.
Its coefficient is a scale dependent coupling at the scale $M$:
$
\frac{1}{g_M}=\frac{1}{g_\Lambda}+\sum_{i=0} \tilde c_{i0} g^n_\Lambda
$
and can be used as an effective expansion parameter.
 One then can reexpress $g_\Lambda$ via
$g_M$ in the other terms of eq.(\ref{L_Mgen}) as well and take the continuum
limit $g_\Lambda\rightarrow 0, \Lambda\rightarrow\infty$ with $\Lambda_{model}$
fixed. Now we introduce
$\beta\equiv 1/g_M$ and $\alpha$ which is analytic function of $g_M$
(typically a small coefficient times $g_M$ or $g_M^2$). Then the effective
lagrangian takes a form:
\begin{equation}
{\cal L}_M[\phi]=\beta_M \left[ {\cal L}_0 + \sum_{i=1}^\infty\sum_{j=1}^\infty
 c_{ij} \alpha^i_M
\frac {1}{M^{2j}} {\cal L}_j(\phi)\right]
\label{L_M}
\end{equation}

An arbitrary physical quantity, like an expectation value of an
operator $O(\phi)$,
can be expanded in $g_M$ (practically, in corresponding $\alpha_M$) and
in $1/M^2$
 (which means in ${\rm (relevant\ \ momentum)}^2/M^2$) as:
\begin{equation}
<O>=\frac{\int_{\phi}O(\phi)  e^{- \int{\cal L}_M}}{\int_{\phi}  e^{
-\int{\cal L}_M}}=
<O>_c+\beta_M \sum_{i=1}^\infty\sum_{j=1}^\infty c_{ij} \alpha^i_M
\frac {1}{M^{2j}} <O{\cal L}_j(\phi)>_c+...
\label{<O>}
\end{equation}
where $<...>_c$ means the connected part of the
expectation value in the coarse grained version
of the original theory and the higher order terms come from the expansion
of the exponent.

2. Now the strong coupling expansion is applied  to calculate these VEVs
assuming that the strong coupling expansion parameter $\beta\equiv1/g_M$
is independent of $\alpha$ in eq.(\ref{<O>}). Note, that
the leading order term in $\alpha$
coincides with   the conventional
"phenomenological" strong coupling model, in which the inverse lattice spacing
 $M$, however, is limited
to values inside the strong - weak applicability window.

The complexity of such a calculation depends on the quantity
and the precision one would like to achieve. Usually strong coupling
series are easy to evaluate to a very high orders. Consequently
the scale $M$ can be
chosen in such a way that $\beta_M$ is just below
the strong coupling expansion radius of
convergence. Simultaneously, the scale $M$ should
be sufficiently high or alternatively the relevant energy scale
sufficiently low so that just a few orders in derivative expansion are
needed to achieve the desired accuracy. This means that the method
 is limited to
low energy quantities only. The loop expansion  for the effective
action coefficients is
probably the most tedious part. An example of such a complexity
estimate is given later.

Rather then describe the general method in more detail, I will proceed
to apply it
to a few simple models in which it is very transparent. I will start with
the simplest possible one:
the Ising chain
\begin{equation}
Z_\Lambda=\sum_{s_x=\pm1} e^{ \frac{1}{g_\Lambda} \sum_x S_x S_{x+1} }
\label{Ising}
\end{equation}
 considered as a $d=1$ QFT. Asymptotic freedom in language of
statistical mechanics means phase transition at zero temperature.
This occurs in d=1 Ising model and generally in d=1 $O(N)$
 symmetric  fixed length spin systems.
In d=1 nearest neighbours interaction
 spin systems the RG transformation can be performed exactly
\cite{kadanoff} and then compared with the
 weak coupling (low temperature) expansion.  This generally
is not available even for d=2 models for which the S - matrix is known.

The exact
"decimation" of the Ising chain
 from the scale $a\equiv 1/\Lambda$ to
$A\equiv1/M$ is especially simple. Not only does the interaction
remain the nearest neighbours one, in addition
 the coarse grained effective action
contains just the original term with a new coefficient \cite{kadanoff}:
$$\frac {1}{g_M}={\rm Arctanh}
\left[{\rm Tanh}^{M/\Lambda}\left(\frac {1}{g_\Lambda}\right)\right]=
$$
\begin{equation}
\frac {1}{g_\Lambda}-\frac{1}{4}
\,{\rm Log} \left(\frac {\Lambda^2}{M^2}\right)+
\frac {1}{6}\frac {\Lambda^2}{M^2}e^{ -4/g_\Lambda}+
O\left(\frac {\Lambda^4}{M^4} e^{ -8/g_\Lambda}\right)
\label{decimation}
\end{equation}
The perturbative expression is obtained by neglecting  exponentially
small pieces:
$1/g_M=1/g_\Lambda-\frac{1}{4} \,{\rm Log} \left(\Lambda^2/M^2\right)$.
This is the usual logarithmic scale dependence,
similar to that in the d=4 YM theories or
the d=2 Gross - Neveu and $\sigma$ models,
corresponding to the following $\beta$ function:
$\beta(g)=-\frac{1}{2} \, g^2$.
Note that there are no contributions to the $\beta$ function
higher then the second order in the coupling.
The corresponding $\Lambda$ parameter $\Lambda_I$ is defined by:
$
\Lambda_I\equiv \Lambda e^{- \int_{\infty}^{g_\Lambda} \frac{1}{\beta (g)}}
=\Lambda  e^{ -2/g_\Lambda}
$
The perturbative expression takes into account only
 the lowest energy configuration, Fig.4a,b, while, as will be shown later,
the exponentially small pieces account for  kink - antikink pairs, Fig 4c,d.
Since the effective action contains only one operator
rather than the usual infinity
of various operators, no derivative expansion is needed.
Now to complete the step 1, we just have to express $g_M$ via $g_\Lambda$
and $\Lambda_I$ and take the limit $\Lambda \rightarrow \infty$.
The result is:
$1/g_M= \frac {1}{4} {\rm Log} \left(M^2/\Lambda_I^2\right)$.

To perform step 2  for
the mass of the single existing excitation one simply exactly solves the
effective model and then uses the exact expression
to generate the strong coupling
series in $\beta_M\equiv 1/g_M$:
\begin{equation}
m/M= - {\rm Log [ Tanh} (\beta_M) ]=-{\rm Log}(\beta_M)+\frac {1}{4} \beta_M^2
-\frac {7}{9} \beta_M^4+ \frac {62}{2835} \beta_M^6+...
\label{strongI}
\end{equation}
The deviation of this  from the exact value,  $m_{exact}/\Lambda_I=2$, for
various $M$ is:
\begin{equation}
\frac {m-m_{exact}}{\Lambda_I}=
\frac {2}{3}\left( \frac {\Lambda_I}{M}\right)^2
+\frac {2}{5}\left( \frac {\Lambda_I}{M}\right)^4+O\left( \left( \frac
{\Lambda_I}{M}\right)^6 \right)
\label{deviationI}
\end{equation}
 The radius of convergence of the strong
coupling series in eq.(\ref{strongI}) is $\beta_{max}=\pi/2$.
At this point $M_{max}=e^\pi\Lambda_I\sim 23.1 \Lambda_I.$
The maximal precision one can achieve is therefore
 $(m_{exact}-m)/m_{exact}\sim 10^{-3}$.

In this very simple example only strong coupling expansion came into full
play. The other two ingredients, namely the weak
 coupling expansion and the derivative expansion, have not.
The weak coupling expansion for $1/g_M$ terminated at the one loop level
and higher loop effects for physical quantities are just
the RG improvement,
while the derivative expansion was not needed.

Now we consider a slightly more complicated solvable
 model in which these two elements
already appear: the O(3) symmetric spin chain:
\begin{equation}
Z_\Lambda=\sum_{{\vec S_x}^2=1}
 e^{\frac {1}{g_\Lambda} \sum_x {\vec S_x} {\vec S_{x+1}}}
\label{ZH}
\end{equation}
This model has a phase transition at zero temperature, although the
asymptotic freedom is powerwise rather then logarithmic:
$\beta(g)=-g+...$ \cite{stanley}.
Using the transfer matrix method, one is  still
able to perform an exact decimation.
The  infinite dimensional transfer matrix
$K({\vec S},{\vec S'})\equiv e^{\frac{1}{g_\Lambda} \sum_x {\vec S} {\vec S'}}$
can be diagonalized in the basis of spherical harmonics \cite{joyce}:
$$ K_\Lambda({\vec S},{\vec S'})=
{\sqrt \frac{\pi g_\Lambda}{2}}\sum_{l=0}^{\infty}
 (2l+1) I_{l+1/2}(1/g_\Lambda)\sum_{m=-l}^{l}4\pi\,
Y_{lm}^\dagger(\theta',\phi')
Y_{lm}(\theta,\phi)= $$
\begin{equation}
\sqrt{\frac {\pi g_\Lambda}{2}}\sum_{l=0}^{\infty}
 (2l+1) I_{l+1/2}(1/g_\Lambda)\,  P_l({\vec S}\,{\vec S'})\equiv
\sum_{l=0}^{\infty}
 (2l+1) F_l^\Lambda(1/g_\Lambda)\, P_l({\vec S}\,{\vec S'})\ \,\,\,\,    \
\end{equation}
$F_l$ are eigenvalues of the matrix.
Here $\theta$ and $\phi$ are the spherical coordinates and $I_{l+1/2}$
and $P_l$ are modified Bessel functions of the first
 kind and the Legendre polinomials respectively.

After the decimation the effective action
still contains only nearest neighbours interactions, but now
 all the powers of
${\vec S} \Box {\vec S}$, where $\Box$ is the
lattice second derivative, are present:
 ${\cal L}_M=const+\frac {1}{g_M}{\vec S} \Box {\vec S}+\sum_{i=1} c_i
({\vec S} \Box {\vec S})^i$.
The coarse  grained
lattice's
transfer matrix is given by the $(\Lambda/M)$s power of $K_\Lambda$. It
means that the eigenvalues are: $F_l^M=(F_l^\Lambda)^{\Lambda/M}$.
Introducing  $\Lambda_H\equiv \Lambda/g_\Lambda$
and taking the continuum limit, one finds the effective
action
\begin{equation}
F_l^M=e^{-\frac{\Lambda_H}{M} \frac {l (l+1)}{2}}.
\label{perf}
\end{equation}
 As in the case of Ising , this in fact defines a perfect action
\cite{hasenfratz}.
 Generally, for a nearest neighbour model
with eigenvalues $F_l$ the spectrum of excitations
is given by
\begin{equation}
m_l/M={\rm Log} \left( \frac {F_l}{F_0}\right)      .
\label{mass}
\end{equation}
Using this one
 can easily check that all the continuum masses, which are simply levels
of the quantum rotator, are correct: $m_l=\frac {l (l+1)}{2}\Lambda_H$.
The coefficients of the first few terms in the derivative expansion of
an action defined by the transfer matrix $F$ are determined by
following equations:
$$z\equiv\sum_{l=0}^\infty
(2 l+1)F_l$$
$$z/g_M=\sum_{l=0}^\infty
(2 l+1)\frac {l (l+1)}{2}F_l$$
\begin{equation}
z\left( \frac{1}{2 g_M^2}+c_1\right )=\sum_{l=0}^\infty
(2 l+1)\frac {3 (l-1) l (l+1) (l+2)}{4!}F_l
\label{coef}
\end{equation}
where $z$ is a physically irrelevant constant.
Taking here the action defined by eq.(\ref{perf})
and using asymptotic form of the Bessel functions \cite{AS},
one finds that the original
${\vec S} \Box{\vec S}$ dimension one term comes with a
coefficient
$
1/g_M=1/\tilde g-1/6-\tilde g/180$,
while that of the next one is:
$
c_1=-1/3 \tilde g +7/90 +11 \tilde g/2700 $. Here $\tilde g\equiv\Lambda_H/M$.

Now the resulting model can be solved exactly.
To leading order in the derivativeand the weak coupling  expansions,
 one gets the known result for the energy gap \cite{stanley}:
$m=-M\, {\rm Log}\left[{\rm Coth}(1/\tilde g)-\tilde g\right]$. This can be
used to generate the strong coupling series.
This  should be compared with the exact
 value $m_{exact}/\Lambda_H=1$.
The difference expanded in inverse powers of $M$ is:
\begin{equation}
\frac {m-m_{exact}}{\Lambda_H}=
\frac {1}{2}\left( \frac {\Lambda_H}{M}\right)
+\frac {1}{3}\left( \frac {\Lambda_H}{M}\right)^2+O\left( \left( \frac
{\Lambda_I}{M}\right)^3 \right)
\label{deviationH}
\end{equation}
 The  strong coupling series radius of convergence is now a bit larger
than that for the Ising chain, $\beta_{max}=\pi$. The
corresponding maximal scale is however very low, $M_{max}= \pi \Lambda_H$,
due to the fast running of the coupling.
In order to obtain better
precision, more orders in the expansion are needed.
 A peculiarity of this model
is that weak coupling corrections are of the same order as the derivative
expansion corrections since the asymptotic freedom is a powerwise one.
So they should be computed together. I calculate here only the
 next to leading correction.

For this, one should take into account the next to leading
order contribution in $\tilde g$ to $1/g_M$ and the
leading contribution
to the coefficient of the four derivative term.
To find the spectrum one should
 reconstruct the eigenvalues $F_l$ from which the mass
is obtained using eq.(\ref{mass}). This can be done using the completeness
relation of the Legendre polynomials \cite{AS}:
 $F_l=\int_{-1}^1 dy e^{1/g_M(y-1)+c_1(y-1)^2}P_l(y)$.
To current order this integral still can be expressed in terms of the error
function,
 but it is more instructive to perform it perturbatively
in $\tilde g$:
$$F_0=\int_{-1}^1 dy e^{1/{\tilde g}(y-1)} \left[1-\frac {1}{6}(y-1)
 - \frac {1}{6\tilde g}(y-1)^2\right]=\tilde g+\frac {5}{6} {\tilde g}^2+...
$$
\begin{equation}
 F_1=\int_{-1}^1 dy y e^{1/{\tilde g}(y-1)} \left[1-\frac {1}{6}(y-1)
 - \frac {1}{6\tilde g}(y-1)^2\right]=\tilde g+\frac {1}{6} {\tilde g}^2+...
\end{equation}
Using the formula for the energy gap eq.(\ref{mass}), one gets the
correct result within a precision of $1/M_{max}^2\sim 0.1$
and the first term in eq.(\ref{deviationH}) disappears. This can be continued
to higher orders.

Now I turn to a two dimensional logarithmically
asymptotically free models which have more similarity to
 four dimensional gauge theories: the Gross - Neveu and the $\sigma$ - models.
I will describe here the former only,
 since the results are similar (which should not be too
surprising, since in many respects the models behave similarly).
The model defined by the lanrangian
\begin{equation}
{\cal L} = i\bar\psi^a\dslash\psi^a + {g\over 2N} (\bar\psi^a\psi^a)^2
\end{equation}
($a=1,...N$) has a rich particle spectrum and
exhibits dynamical (discrete) chiral symmetry breaking.
Although the exact S -  matrix of the theory is known \cite{zz},
in order to compare the expansions with exact results at every step
I will consider the model in the large number of flavours limit only.
This time, however, the $1/M^2$ expansion to
an arbitrary order will be calculated.
The first step is the perturbative integration of the
high frequency modes. It will become clear shortly that only
effective potential
\begin{equation}
V(\bar\psi\psi)=
\sum_{n=1}^{\infty}c_n M^2 \left( \frac{1}{M}\bar\psi\psi\right)^n
\end{equation}
will be needed to calculate the fermion's mass in the large $N$ limit
(of course, even within the large {\it N} approximation there
might appear additional irrelevant
operator, those containing derivatives, like $\bar\psi\dslash\Box \psi$ or
$\bar\psi\psi\Box\bar\psi\psi$).
Perturbatively,  only the one particle
irreducible diagrams shown on Fig.1 contribute.


%
%
%
%
%
%
%
%
The internal loops carrying momenta between $M$
 and $\Lambda$  only are integrated and external momenta are all
 zero\footnote{We use the sharp
momentum cutoff regularization which at this level does not lead to any
ambiguities although the calculation (together with analogous calculation in
the nonlinear $\sigma$ - model) can be done,
for example, on  the
lattice. Although this is
very helpful to actually perform strong coupling
 expansion for finite $N$ \cite{rz}, it
unnecessarily complicates the large $N$ calculation.}.
Other 1PI diagrams vanish due to (unbroken at this stage) chiral symmetry
or are of lower order in $1/N$.
The first, the third and similar chain diagrams determine
the  relevant  dimension two term renormalization:
\begin{equation}
\frac{1}{g_M}=\frac{1}{g_\Lambda}
-\frac{1}{2\pi} {\rm Log} \left(\frac{\Lambda^2}{M^2}\right),
\end{equation}
while the second is the coefficient of the dimension four irrelevant operator:
\begin{equation}
c_2=g_\Lambda^4\frac{1}{8\pi}\left(\frac{1}{M^2}-\frac{1}{\Lambda^2}\right)
\end{equation}
Reexpressed via $\Lambda_{GN}=\Lambda  e^{-\pi/g_\Lambda}$ and at the continuum
limit $\Lambda\rightarrow\infty$, they become: $c_1=g_M/2$ and
$c_2=g_M^4\frac{1}{8\pi}\frac{1}{M^2}$.

At first,  I neglect the
$1/M^2$ corrections. The mass should be evaluated
using the strong coupling expansion. Instead, I will
find it exactly and then expand in $\beta_M$. The large $N$
 dynamically generated mass $m_0$ is the solution of the gap equation:
\begin{equation}
\frac{1}{g_M}=\frac{1}{\pi}  \int_0^M \frac {p\,  dp}{p^2+m_0^2}=
\frac{1}{2\pi}{\rm  Log} \left(\frac {M^2+m_0^2}{m_0^2}\right)
\label{gapGN}
\end{equation}
In terms of $\Lambda_{GN}$ it is
\begin{equation}
\frac{m_0^2}{\Lambda_{GN}^2}=\frac{1}{1-(\Lambda_{GN}/M)^2}= 1+
\frac {\Lambda^2_{GN}}{M^2}-\left(\frac {\Lambda^2_{GN}}{M^2}\right)^2+...
\label{mGN}
\end{equation}
and should be compared with the exact value $m_{exact}/\Lambda_{GN}=1$.

Now I will show that the leading $1/M^2$ correction eliminates
the second term in eq.(\ref{mGN}) thus improving the precision to O($1/M^4$).
The fermion propagator to this order in $1/M^2$ is
\begin{equation}
G(p)=G_0(p)-c_2\frac{1}{M^2}<\bar\psi(p)\psi(-p)\int_x(\bar\psi\psi)^4>_c
\end{equation}
where $<...>_c$ denotes  connected $1/N$ evaluated
 contributions shown on Fig.2.

All the $1/N$ diagrams (for the Feynman
rules, see, for example, \cite{rwp}) are
evaluated with mass $m_0$.
Note also that the other dimension four operators like
$\bar\psi\dslash\Box\psi$ do not contribute; they all vanish in the
large $N$ limit (see Fig.1).
The result for the mass is (there is no
wave function renormalization at large $N$) $m_1^2/\Lambda_{GN}^2=1+O(1/M^4)$,
as advertized.
Higher orders in derivative expansion gradually help to recover the precise
result.

%
%
%
%
%
%
%
%
%

The radius of convergence of the strong coupling series is evident from the
exact solution eq.(\ref{gapGN}) for $m_0^2=\frac{M^2}{e^{2\pi\beta_M}-1}$:
$\beta_{max}=1/g_{min}=1$.
Consequently, the maximal possible value of the scale $M$ is then
$M_{max}= e^{\pi}\Lambda_{GN} \sim
23.1\Lambda_{GN}$.
Now we would like to estimate the practicaly achievable
precision of such a scheme. Suppose we would like to
calculate to order $n_s$ in strong coupling (so that we neglect
the order $\beta^{(n_s+1)}$), to order $n_w$ in weak coupling and to
order $2n_d$ in derivative expansion. Then the precision of the mass,
$\Delta\equiv (m-m_{exact})/m_{exact}$,
is given by
$$\Delta=min_M [\Delta(M)]$$
\begin{equation}
\Delta(M)=max \left[\left(\frac
{\beta}{\beta_{max}}\right)^{(n_s+1)},\alpha^{(n_w+1)},\left(\frac
{m^2}{M^2}\right)^{(n_d+1)}\right]
\label{deviationestimate}
\end{equation}
For example, if $n_s=10$, $n_w=1$ and $n_d=1$ one obtains $\Delta=0.045$
for $M=10.7\Lambda_{GN}$ and $\beta=0.75$. If, instead of $n_w=1$,
we are able to calculate $n_w=2$ (no change in $n_s$ and $n_d$), we improve to
achieve a precision of $\Delta=0.015$
at $\beta=0.68$, $M=8.4 \Lambda_{GN}$. In the first example out of three
quantities in eq.(\ref{deviationestimate}) the weak and the strong coupling
expansion determined the optimal $M$, while derivative expansion was not
critical. In the second example the derivative and the strong
coupling determined the optimal $M$, while weak coupling was not critical.


The reader might wonder about the following dynamical question.
What happened to
spontaneous chiral symmetry breaking during this calculation?
On the one hand, we used the weak coupling perturbation theory that does not
know about the symmetry breaking to calculate the effective potential and only
later introduced the strong coupling expansion.
The strong coupling expansion (actually an exact solution of the perturbative
effective
action) recovered the full effect of the chiral symmetry breaking. We would
expect that, at least some imprecision would be introduced by the first step.
What happened here in the large $N$ limit is that the effects
 of the dynamical symmetry breaking
are felt by the effective potential only when the scale $M$ drops below
$\Lambda_{GN}$. This question is interesting in its own right, so
 I calculated the exact large $N$
effective potential using methods first applied in \cite{ma} to  bosonic
models.

Here only the result
is presented:
\begin{equation}
V(u)=\frac{1}{4\pi}\left[ 2(\Lambda^2-M^2)+w^2-
M^2 {\rm Log}\left(\frac{M^2}{\Lambda_{GN}^2}\right)\right]+
\frac{1}{2}\left(\frac {M^2}{w}-w\right)u
\end{equation}
where $u$ denotes $\bar\psi\psi$ and $w(u)$ is a solution of
\begin{equation}
u=\frac {w}{2\pi}{\rm Log}\left(\frac{M^2+w^2}{\Lambda_{GN}^2}\right).
\end{equation}
For few values of $M$ it is given on Fig. 3.

 Differentiation the potential at
the origin for $M>\Lambda_{GN}$, one recovers the coefficients $c_i$
to any order.
One can see in Fig. 3 how the shape of the
effective potential changes as more modes are being integrated out.
Only when the modes below dynamically generated mass are integrated
out will there be nonperturbative effects. For example, the slope at
the origin of the
potential for all the values $M>\Lambda_{GN}$ is exactly zero,
but below $\Lambda_{GN}$ is starts rising till it reaches
the value of the dynamically generated fermion mass at $M=0$.

Let us return now to the double expansion.  Notice that, since
even at high energies there are nonperturbative effects, the
 use of the perturbation theory, even just for
integration of high energy modes, should  generally
introduce a systematic error.
We have seen this in the Ising chain case. What is the nature of
these errors and can they be reduced? At least in the case of the Ising chain,
the answer is that they can be systematically removed if one takes into
account instantons. Indeed, I will show now
that the difference between the perturbative
effective action and the exact one  eq.(\ref{decimation})
is caused by the multi instanton
effects.
The Ising chain is a limiting version of the quantum mechanical
double well potential. So it comes as no surprise that they
appear \cite{P}.

 Suppose we are decimating from some small
 scale $a=1/\Lambda$
to a much larger scale $A=1/M\equiv b\, a$.
When the two adjacent spins on the $M$ lattice are parallel, there is just one
configuration, Fig. 4a, of minimal energy. Therefore
at low temperature,
the sum
$\sum_{S_1=\pm 1}...\sum_{S_K=\pm 1}
 e^{\frac{1}{g_\Lambda}\sum_xS_xS_{x+1}}$
can be approximated by $Z^0_{\uparrow\uparrow}\sim e^{\frac {b}{g_\Lambda}}$.
If the spins are antiparallel, Fig. 4b, the lowest energy state  with
one flip is $b$ times degenerate and the sum within the "steepest descent"
is $Z^0_{\uparrow\downarrow}\sim b \,  e^{\frac{(b-2)}{g_\Lambda}}$. The
coefficient in the effective
lagrangian therefore is given by
$e^{2/g_M}=Z^0_{\uparrow\uparrow}/Z^0_{\uparrow\downarrow}$ and coincides
with
the first two terms in eq.(\ref{decimation}).

 Now let us take into account
 one flip - antiflip. For the parallel spins, Fig 4c, one adds
$Z^1_{\uparrow\uparrow}\sim b^2/2 \, e^{\frac {(b-4)}{g_\Lambda}}$,
while for the antiparallel spins, Fig. 4d, the
contribution is $Z^1_{\uparrow\downarrow}\sim b^3/6 \, e^{\frac
{(b-6)}{g_\Lambda}}$. Now
 the third term in the expansion of effective coupling in small
exponentials of the bare coupling, eq.(\ref{decimation}) is correctly
 reproduced.
Therefore if one is to improve on the precision achieved earlier, one
 simply has to take instantons into account.
Note, that the instanton calculus is performed for small coupling and no
problem with overlapping instantons should arise.
Logarithmically asymptotically free models are special in that
contributions of instantons to the effective action is actually of the same
order in $1/M^2$ as the perturbative contributions from irrelevant operators.
Indeed, the exponential factors in effective action in terms of $\Lambda_I$
are simply inverse powers of $M$. Consecuently, the precision obtained earlier
$m^2/M_{max}^2\sim 10^{-3}$ can be pushed lower
by including one kink - antikink
effects to  $m^4/M_{max}^4\sim 10^{-6}$.

In other logarithmically asymptotically free models like
the $\sigma$ - model
or $d=4$ nonabelian gauge theories similar
 contributions to the derivative expansion are expected.
They might not show up though in the large $N$ limit, as the
case of the Gross - Neveu and the $\sigma$ models show. Currently this
contribution in the
$O(3)$ symmetric $\sigma$ model is being investigated \cite{rz}.

To summarize,  the double strong - weak expansion method is
proposed for a quantitative study of low energy phenomena in
asymptotically free theories. One integrates the high frequency modes using
weak coupling expansion (and to further increase the precision, instanton
calculus) and then the effective theory is treated using
the strong coupling expansion.  It was  applied to some
solvable low dimensional asymptotically free theories and is able to reproduce
the exact results.
The application of the method to realistic four dimensional
gauge theories should be addressed next.

Discussions with T. Bhattacharya, A. Kovner,  V. Kushnir,
A. Speliotopoulos
and V. Zhytnikov are greatly appreciated.
I  acknowledge the support of
National Science Council of ROC, grant
NSC-83-0208-M-001-011.


\newpage

\begin{thebibliography}{99}

\bibitem{kreutz} M. Creutz, {\it "Quarks, gluons and lattices."}.
Cambridge University Press, New York (1983).
\bibitem{kogut} J. Kogut, D.K. Sinclair and L. Susskind, {\it Nucl. Phys.}
 {\bf B114}, 199 (1976); G. Munster,  {\it Nucl. Phys.}
 {\bf B190} , 439 (1981); K. Seo,  {\it Nucl. Phys.}
 {\bf B209}, 200 (1982); J. Smit  {\it Nucl. Phys.}
 {\bf B206}, 309 (1982; H. Kluberg-Stern, A. Morel and B. Peterson,
 {\it Nucl. Phys.}
 {\bf B215}, 527 (1983); K. Kimura,  {\it Nucl. Phys.}
 {\bf B246}, 143 (1984);  J. Hoek and J. Smit,
 {\it Nucl. Phys.}
 {\bf B263}, 129 (1986).
\bibitem{davis} M.G. Mitchard, A.C. Davis and J.A. Gracey,
{\it "Realization in the chiral Gross-Neveu
model"}, preprint DAMTP/88-23, (1988); M. Brunelli and  M.Gomes,
 {\it  Z.Phys.} {\bf C42},649 (1989).
\bibitem{elitzur}
J. Shigemitsu and S. Elitzur, {\it Phys. Rev.} {\bf D14},
1988 (1976).
R. Musto et al.  {\it Nucl. Phys.} {\bf B210}, 263
(1982); P. Butera, M. Comi and G. Marchesini,  {\it Nucl. Phys.} {\bf B300}, 1
(1988);
M. Luscher and P. Weisz,  {\it Nucl. Phys.} {\bf B300}, 325
(1988).
\bibitem{polchinski}K. Wilson and J. Kogut, {\it Phys. Rep.} {\bf 12C}, 75
 (1974) ;
 J. Polchinski, {\it Nucl.Phys.} {\bf B231}, 269 (1984);
R.S. Ball and R.S. Thorne, {\it Renormalizability of effective
scalar field theory"}, hep-th@xxx.lanl.gov - 9310042 (1994)
and references therein.
\bibitem{kadanoff} L.P. Kadanoff, {\it Ann. Phys. (NY)} {\bf 100},359 (1976).
\bibitem{zz} A. B. Zamolodchikov and A. B. Zamolodchikov {\it Ann. Phys} {\bf
120} , 239 (1979).
\bibitem{joyce} G.S Joyce, {\it Phys. Rev.} {\bf 155},
478 (1967).
\bibitem{stanley} H.E. Stanley,  {\it Phys. Rev.} {\bf 179},
570 (1969); D. Fisher and D.R. Nelson, {\it Phys. Rev.} {\bf B16}, 2300 (1977).
The last paper (like few other papers)
 contains an attempt to simply glue the strong and weak
 coupling series for certain
quantities calculable in both weak and strong couplings
 without use of the Pade approximants.
\bibitem{AS} M.Abramowitz and I.A. Stegun,
{\it "Handbook of Mathematical Functions"}, Dover Publishers, New York (1964).
\bibitem{hasenfratz}
P. Hasenfratz and F. Niedermayer,  {\it Nucl. Phys.} {\bf B414}, 785 (1994).
\bibitem{rz} B. Rosenstein and V. Zhytnikov, work in progress.
\bibitem{rwp} See, for example, B. Rosenstein, B. Warr and S.H. Park,
{\it Phys. Rep.} {\bf 205}, 59 (1991).
\bibitem{ma} S.K. Ma,  {\it Rev. Mod. Phys.} {\bf 45}, 589 (1973).
\bibitem{P} A. M. Polyakov, {\it "Gauge Fields and Strings"}, Harwood Acad.
Publishers, New York (1987).
\end{thebibliography}
\end{document}